\documentclass{article}       
\usepackage{e}		      
\begin{document}
\branch{C}   
\DOI{0005}			  
\idline{C}{5, 1--12}{1}	  
\editorial{12 Jan 2001}{24 Apr 2001}{ 30 Apr 2001}{}
		  
\title{
Next-to-leading order mass effects in QCD Compton
process of polarized DIS
}
\author{
I.Akushevich\inst{1}
\thanks{on leave
of absence from the National
Center of Particle and High Energy Physics,
220040 Minsk, Belarus}
\and A.Ilyichev\inst{2} \and N.Shumeiko\inst{2}
}
\institute{
North Carolina Central University, Durham, NC 27707, USA
and TJNAF, Newport News, VA 23606, USA
\and
National Center of Particle and High Energy Physics,
220040 Minsk, Belarus, e-mail: ily@hep.by}
\PACS{not given}
\maketitle
\begin{abstract}
The method originally developed for the exact calculations in QED
theory is applied for the calculation of NLO effects in QCD Compton
processes. QCD corrections to the structure functions and sum
rules are obtained. Different interpretations of the NLO effects due to
finite quark mass are discussed.
\end{abstract}

\section{Introduction}

A lot of articles dedicated to NLO QCD corrections to polarized
DIS have been published. Calculations in most of them
are performed in an assumption that the quark mass
is equal to zero (see \cite{AGL,Kod1,Kod2,Kod3,AEL,PR,GRV} and
review \cite{QCD}).
However there are some works when authors estimate
a finite quark mass effect for both unpolarized \cite{NE1}
and polarized structure functions
\cite{NE2,AG2,TER,TERV,TERM,NE3,BT}. The
main difference of these two approaches is the	method of
tending the quark mass to zero and the procedure of integration
of the squared matrix element
over the phase space of emitted gluon. In the massless approsch
a fermion
mass is equal to zero {\it before } the integration while in the
massive one this quantity
goes to zero {\it after} the integration and
survives only in LO terms.

Notice that QCD and QED radiative corrections (RC) having different
origins
possess some common features on the one-loop level.
If our consideration is restricted to so-called QCD Compton
process then both corrections should be described by the
identical sets of Feynman graphs. The transition from the strong
radiative effects to the electromagnetic ones could be performed
by the following replacement:
\begin{equation}
\frac 4 3 \alpha _s \rightarrow e_q^2\alpha _{QED}.
\end{equation}

In the present paper we investigate QCD Compton
process for polarized nucleon target by the method
developed for exact calculations \cite {BS,ASh,P20,LRC,ABK,BK}
in QED theory for massive
polarized fermions but never used  for QCD.
We estimate the influence of the finite quark mass effects on the
value of the polarized
sum rules and discuss some different interpretations of the first
moment of the polarized structure function $g_1$.

The present paper is organized in the following way.
The method of calculation is described in Section 2.
 The explicite
expressions for the QCD-improved structure functions as well as their
first moments can be found in Section 3.
Then some conclusions are made in Section 4. We  also discuss there
different points of view how the NLO finite quark mass effects
influence the first moment of $g_1$.
Details of the integration procedure are
presented in Appendix.

\section{Method of Calculation.}

The cross section of polarized lepton-nucleon DIS
\begin{equation}
l(k_1, \xi)+N(p,\eta) \rightarrow l(k_2)+ X
\end {equation}
on the level of the one-photon exchange can be presented as
a convolution of the well-known lepton tensor:
\begin{equation}
L_{\mu \nu}=\frac 12 Sp \; \gamma _{\mu }(\hat
{k}_1+m)(1+\gamma_5 \hat {\xi})\gamma_{\nu}(\hat {k}_2+m),
\label {l1}
\end {equation}
and the hadronic one:
\begin{eqnarray}
W_{\mu \nu}=&&-g_{\mu \nu}F_1
+\frac {p_{\mu }p_{\nu}}{pq}F_2
\nonumber \\&&
+\frac{iM}{pq}\epsilon_{\mu \nu \lambda \sigma }q^{\lambda }
({\eta }^{\sigma }g_1+
(
{\eta}^{\sigma }-
\frac {\eta q}{pq}
p^{\sigma })
g_2).
\label{w}
\end{eqnarray}
Here $p$  and $\eta $ are an initial 4-momentum of the target and
its polarization vector respectively and $q$ is a 4-momentum of the
virtual
photon ($p^2=M^2$, $q^2=-Q^2$).

On the Born level in the frame of the naive parton model the
structure
functions contributed to the hadronic tensor $W^0_{\mu \nu }$ have the
following form:
\begin{eqnarray}
F_1^0(x)=&&
\frac 12 \sum _q e_q^2f_q(x),
\nonumber \\
F_2^0(x)=&&
x \sum _q e_q^2f_q(x),
\nonumber \\
g_1^0(x)=&&\frac 12 \sum _q e_q^2\Delta f_q(x),
\nonumber \\
g_2^0(x)=&&0.
\label{0sf}
\end {eqnarray}
Here $f_q(x)$ and $\Delta f_q(x)$ are unpolarized and polarized parton
distributions respectively{\footnote{Here and later the
independence of the polarized parton
distribution on the polarization vector is assumed. See discussion in
\cite{AEL}.}, and $x=Q^2/2pq$ is the standard Bjorken variable.

The one-loop lowest-order RC to the hadronic current consists of
two parts with contributions calculated in different ways:
\begin{equation}
W^{1-loop}_{\mu \nu}=
W^V_{\mu \nu}+
W^R_{\mu \nu}.
\label{1loop}
\end{equation}
The first part of the one-loop RC to the hadronic current comes
from the gluon exchange and it will be described later.
The second one corresponds to the gluon emission and requires the
integration over its phase space:
\begin{eqnarray}
W^R_{\mu \nu} &=&
\frac {\alpha_s}{12\pi ^2}
\sum _q e_q^2  \int
\frac {d^3k}{k_0}\frac 1 {(p_{2q}-p_{1q})^2}
\nonumber \\&&
\times[w^{Ru}_{\mu \nu}f _q(x/z)+w^{Rp}_{\mu \nu}\Delta f
_q(x/z)],
\label {wra}
\end{eqnarray}
where
\begin{eqnarray}
w^{Ru}_{\mu \nu} &=&
Sp \; \Gamma _{\mu \alpha}
(\hat {p}_{1q}+m_q)
\bar{\Gamma } _{\alpha \nu}(\hat {p}_{2q}+m_q),
\nonumber \\
w^{Rp}_{\mu \nu} &=&
Sp \; \Gamma _{\mu \alpha}
(\hat {p}_{1q}+m_q)
\gamma_5 \hat\eta
\bar{\Gamma } _{\alpha \nu}(\hat {p}_{2q}+m_q),
\label {wra2}
\end {eqnarray}
and $p_{1q}$ ($p_{2q}$) is an initial (final) 4-momentum of the
quark
($p_{1q}^2=p_{2q}^2=m_q^2$).

Quantities $\Gamma _{\mu \alpha}$ and $\bar{\Gamma} _{\alpha \nu
}$ are defined as
\begin{eqnarray}
\Gamma _{\mu \alpha}=&&
2\Omega _q^{\alpha }
\gamma_{\mu}-
\frac {\gamma_{\mu}\hat{k}\gamma_{\alpha}}{v_q}-
\frac{\gamma_{\alpha}\hat{k}\gamma_{\mu}}{u_q}
,
\nonumber \\
\bar{\Gamma} _{\alpha \nu}=&&
2\Omega _q^{\alpha }
\gamma_{\nu}
-\frac{\gamma_{\alpha}\hat{k}\gamma_{\nu}}{v_q}
-\frac {\gamma_{\nu}\hat{k}\gamma_{\alpha}}{u_q}
,
\label{g12}
\end{eqnarray}
where
\begin{equation}
\Omega _q=
\frac {p_{1q}}{v_q}- \frac {p_{2q}}{u_q},
\end{equation}
and
$v_q=2kp_{1q}$, $u_q=2kp_{2q}$.
A variable $z=Q^2/(Q^2+u_q)$ is a standard one used for
the description of QCD effects.

Within the requirements of the naive parton model in (\ref {wra}) we
assume that the quark mass depends on the variable $z$ as
\begin{equation}
m_q=x/z M
\label{mq}
\end{equation}
for bremsstrahlung process and
\begin{equation}
m_0=m_q(z=1)=xM
\label{m0}
\end{equation}
for non-radiative one.
Notice that the factor before $M$ in these formulae coincides
with the argument of the
corresponding parton distribution.

 Both contributions in right hand side in (\ref{1loop}) include
the infrared divergences
to be carefully considered for canceling.
Similar to QED we use an identity:
\begin{equation}
W^R_{\mu \nu }=
W^R_{\mu \nu }-W^{IR}_{\mu \nu}+W^{IR}_{\mu \nu }=
W^{F}_{\mu \nu }+W^{IR}_{\mu \nu }.
\end{equation}

Here an infrared part can be written in the following way:
\begin{equation}
W^{IR}_{\mu \nu }=\frac 4 3 \frac {\alpha_s }{\pi }\sum _q\delta
^{IR}W^{0q}_{\mu \nu },
\label{ikr}
\end{equation}
where $W^{0q}_{\mu \nu}$ is a contribution of $q$-quark to  the hadronic
tensor on the Born level and
$\delta ^{IR} $ should be decomposed in a sum of a soft and hard parts:
\begin{eqnarray}
\delta^{IR}=&&
\frac 1 {\pi }\int \frac {d^3k}{k_0}F^{IR}=
\frac 1 {\pi }\int \frac {d^3k}{k_0}F^{IR}\theta (\epsilon -k_0)
\nonumber \\&&
+\frac 1 {\pi }\int \frac {d^3k}{k_0}F^{IR}\theta (k_0-\epsilon)
=\delta ^{IR}_{soft}+\delta ^{IR}_{hard}.
\label{sh}
\end{eqnarray}
Here $k_0$ is the energy of the emitted gluon,
$\epsilon$ is an infinitesimal parameter of the separation and
\begin{equation}
F^{IR}=\frac {(Q^2+u_q) } {(Q^2+u_q-v_q)}
\left (\frac {Q^2+u_q}{u_qv_q}-\frac {m_q^2}{u_q^2}-\frac
{m_q^2}{v_q^2}\right ).
\label{fir}
\end{equation}

Notice that there is some arbitrariness in the definition of $F^{IR}$:
the asymptotic expression ($k \rightarrow 0$) is fixed.
We choose it like (\ref{fir}) to write the QCD-improved structure
functions
(see formulae (\ref{qcdsf})) in the simplest form. Recall that the
quark mass in  (\ref{fir}) is defined by (\ref{mq}).

The way of calculation of typical integrals in
the soft and hard parts of $\delta ^{IR}$ is widely
discussed in \cite {ABK,BK}.
After explicit integration the sum of the soft
\begin{equation}
\delta ^{IR}_{soft}=2
(l_0-1)
\left(P^{IR}+\log \frac {2\epsilon}{\mu}\right)
-l_0^2+l_0+1
-\pi ^2/6
\end{equation}
and the hard parts
\begin{equation}
\delta^{IR}_{hard}=2
(l_0-1)
\log \frac {m_0}{2\epsilon}
+\frac 32l_0^2+l_0l_v-\frac
12 l_v^2-l_0
-l_v-\pi ^2/6
\end{equation}
gives the result independent on the parameter $\epsilon$.
Here
\begin{equation}
l_0=\log \frac{Q^2}{m_0^2},\qquad
l_v=\log \frac {1-x}x.
\end{equation}
The pole term which corresponds to the infrared divergence
is contained in
\begin{equation}
P^{IR}=\frac 1{n-4}+\frac 12\gamma _E +\log \frac 1{2{\sqrt \pi
}}.
\end{equation}
 The arbitrary parameter $\mu $ has a dimension of a mass and
$\gamma _E$ is the Euler constant.

To extract some information about QCD contributions to the
polarized structure functions $g_1$ and $g_2$ an integration in $W^F_
{\mu \nu}$ over the gluon phase space should be performed without any
assumptions about the polarization vector $\eta $. So the technique of
tensor integration has to be applied in this case. All exact
expressions for tensor, vector and scalar integrations can be
found in the appendix. The ultrarelativistic limit $m_q
\rightarrow 0$ can be performed in a straightforward way, with
the exception of two terms: $f(x)/\tau$, $u_qf(x/z)/\tau ^2$.  The
first one including $1/{\tau }$ appears only in the substracting
parts of $W^F_{\mu \nu}$. Since the corresponding parton distributions
do not depend on $u_q$ it can be integrated over $u_q$ analytically.
The second  one $u_q/{\tau }^2$  comes	from $W^R_{\nu \mu}$.
The ultrarelativistic limit  in  $W^R_{\nu \mu}$ can be applied
after the transformation
\begin{equation}
\int du_q \frac {u_q}{\tau ^2}f(x/z)=
\int du_q \frac {f(x/z)-f(x)}{u_q}
+f(x)\int du_q \frac {u_q} {\tau ^2}
\label{tau}
\end{equation}
and the analytical integration of the last term.
Here $f$ is any (either unpolarized or polarized) parton distribution.
The first term in the right hand side of (\ref{tau}) gives the
contribution to $W^F_{\mu \nu}$. The second one together with the result
coming from the integration $1/\tau$ over $u_q$
gives  an additional factorized correction $\delta ^m $:
\begin{equation}
\delta ^m=-\frac 14(3l_0+3l_v+1).
\end{equation}
An explicit expression for
the gluon exchange contributions reads:
\begin{equation}
W_{\mu \nu}^V=\frac 43 \frac {\alpha _s}{\pi}\sum _q\delta ^V
W_{\mu \nu}^{0q}+W_{\mu \nu}^{AMM},
\end{equation}
where
\begin{equation}
\delta ^V=-2\left ( P^{IR}+\log \frac {m_0}{\mu} \right)(l_0-1)
-\frac 12 l_0^2+\frac 32 l_0 -2+\frac{\pi ^2}6
\end{equation}
is a vertex correction part factorizing before
$W_{\mu \nu}^{0q}$, and
\begin{equation}
W_{\mu \nu}^{AMM}=\frac {\alpha _s}{6\pi}
l_0\sum _q
\frac{iM}{pq}\epsilon_{\mu \nu \lambda \sigma }q^{\lambda }
\Bigl [
{\eta}^{\sigma }-
\frac {\eta q}{pq}
p^{\sigma }\Bigr ]
\sum _q \Delta f_q(x)
\end{equation}
is the quark anomalous magnetic moment.

The infrared terms $P^{IR}$, parameters $\mu$, $\epsilon$  and
the squa\-red
logarithms containing mass singularities are canceled in the sum
\begin{eqnarray}
\delta ^{IR}+\delta ^m+\delta ^V
=&&\frac  12
\left(2l_0l_v-l_v^2+\frac 32 l_0
-\frac 72 l_v-\frac 52
-\frac {\pi^2}{3}
\right) \nonumber\\
=&&\frac 12\delta _q
\label{delq}
\end{eqnarray}
that is identical to (C.17) from \cite{BBC}.

 Since the result of the analytical integration has the same tensor
structure as the usual hadronic tensor in polarized DIS, the coefficients
before the corresponding tensor structures (like $g_{\mu\nu}$, $p_\mu
p_\nu$ ... ) in the sum $W^V_{\mu \nu }+W^R_{\mu \nu}$ can be interpreted
as one-loop QCD contributions to the corresponding structure functions.

\section{QCD-improved structure functions}

In the last section we obtained one-loop correction to the hadronic
tensor and structure functions. The explicit result for them is:
\begin{eqnarray}
\label{qcdsf}
F_1(x,Q^2)=&&
\frac 1{2x}[F_2(x,Q^2)-F_L(x,Q^2)],\;\;
\nonumber \\
F_2(x,Q^2)=&&
x \sum _q e_q^2f_q(x,Q^2),
\nonumber \\
F_L(x, Q^2)=&&
\frac {4\alpha _s}{3\pi}x\sum _q e_q^2\int
\limits_x^1dzf_q(x/z),\;\;
\nonumber \\
g_1(x,Q^2)=&&
\frac 12 \sum _q e_q^2\Delta f_q(x,Q^2),
\nonumber \\
g_2(x,Q^2)=&&
\frac {\alpha _s}{6\pi}\sum _q e_q^2
\biggl\{
(1-2l_q-\log (1-x)
)\Delta f _q(x)
\nonumber \\&&
+\int \limits ^1 _xdz
\Bigl [ (4l_q
-4\log z(1-z)
-12
\nonumber \\&&
-\frac 1{(1-z)})\Delta f_q(x/z)
+\frac{\Delta f_q(x)}{(1-z)}
\Bigl ]
\biggl\}.
\end {eqnarray}
The $Q^2$-dependent unpolarized and polarized parton distributions
are defined as
\begin{eqnarray}
\label{qcdpd}
f_q(x, Q^2)=
&&\displaystyle
(1+\frac {2\alpha _s}{3\pi} \delta _q)
f_q(x)
+\frac {2\alpha _s}{3\pi} \int\limits_x^1 \frac{dz}{z}
\Bigl[
\Bigl(\frac {1+z^2}{1-z}
\nonumber\\&&
\!\!\!\!\!\!\!\!\!\!\!\!\!\!\!\!\times(l_q-\log z(1-z))
-\frac 72 \frac 1{1-z}
+3z
+4\Bigr)f_q\bigl(\frac{x}{z}\bigr)
\nonumber \\&&
\!\!\!\!\!\!\!\!\!\!\!\!\!\!\!\!-\frac 2{1-z}\left(l_0+\log \frac{z}{1-z}
-\frac 74\right)f_q(x)
\Bigl],
\nonumber \\
[0.5cm]
\Delta f_q(x,Q^2)=&&
(1+\frac {2\alpha _s}{3\pi} \delta _q)
\Delta f_q(x)
+ \frac {2\alpha _s}{3\pi} \int\limits_x^1 \frac{dz}{z}
\Bigl[\bigl(\frac{1+z^2}{1-z}
\nonumber \\&&\displaystyle
\!\!\!\!\!\!\!\!\!\!\!\!\!\!\!\!\times(l_q-\log z(1-z))
-\frac 72 \frac 1{1-z}
+4z
+1\Bigr)\Delta f_q\bigl( \frac{x}{z}\bigr)
\nonumber \\&&
\!\!\!\!\!\!\!\!\!\!\!\!\!\!\!\!-\frac 2{1-z}\left(l_0+\log \frac{z}{1-z}
-\frac 74\right)\Delta f _q(x)
\Bigl],
\end{eqnarray}
where $l_q=\log Q^2/m_q^2$ and $\delta _q $ is defined by (\ref{delq}).

Notice that our formulae
(\ref{qcdsf},\ref{qcdpd}) are in an
agreement with	ones obtained earlier. The unpolarized structure
functions coincide with (2.24,2.49) from \cite{NE1}, $g_1(x,Q^2)$
corresponds to the expression (13) in \cite{NE2}. At last
$g_2(x,Q^2)$
can be compared with the sum (18) and (19) from \cite{AG2}.

Let us calculate the difference between QCD-impro\-ved parton
distributions found in both of the discussed sche\-mes:
\begin{eqnarray}
\frac {3\pi }{4\alpha _s}(
f_q^{m_q= 0}(x, Q^2)-f_q^{m_q\ne 0}(x, Q^2))=
\Bigl[(P^{IR}
\nonumber \\
+\log \frac {m_0 }{\mu })(\frac 32+2l_v)
-l_v^2
-l_v+1\Bigl]f_q(x)
\nonumber \\
+\int \limits_x^1 \frac {dz}z
\Bigl[\Big(\frac {1+z^2}{1-z}
(P^{IR}
+\log \frac {m_q }{\mu }+
\log(1-z))
\nonumber \\
+\frac 1{1-z}
-\frac 12z
-\frac 12\Big)f_q\bigl(\frac{x}{z}\bigr)
\nonumber \\
-\frac 2{1-z}\Big(P^{IR}+\log \frac {m_0 }{\mu }+\log \frac
{1-z}z+\frac 12\Big)f_q(x)
\Bigl ],
\nonumber \\
\frac {3\pi }{4\alpha _s}(
\Delta f_q^{m_q= 0}(x, Q^2)-\Delta f_q^{m_q\ne 0}(x, Q^2))=
\Bigl [(P^{IR}
\nonumber \\
+\log \frac {m_0 }{\mu })(\frac 32+2l_v)
-l_v^2
-l_v+1\Bigl ]\Delta f_q(x)
\nonumber \\
+\int \limits_x^1 \frac {dz}z
\Bigl[\Big(\frac {1+z^2}{1-z}
(P^{IR}
+\log \frac {m_q }{\mu }+
\log(1-z))
\nonumber \\
+\frac 1{1-z}
-\frac32z
+\frac 12\Big)\Delta f _q(\frac xz)
\nonumber \\
-\frac 2{1-z}\Big(P^{IR}+\log \frac {m_0 }{\mu }+\log \frac
{1-z}z+\frac 12\Big)\Delta f _q(x)
\Bigl ].
\label{diff}
\end{eqnarray}
It can be seen from (\ref{diff}) that the partonic distributions within
the two schemes have the same structure and all leading
logarithmic terms are canceled after the following
replacement:
\begin{equation}
\log  m_{q,0} \rightarrow
\log \mu -P^{IR}.
\end{equation}
The standard QCD results include infinite terms $P^{IR}$ (see
ref.\cite{PR}) that corresponds to quark mass singularity terms in our
formulae. These divergent and scheme dependent terms in the
classical results are rejected by the additional procedure of the
renormalization of the parton distributions (see ref.\cite{QCD}
for details). Contrary, our formulae (\ref{qcdpd}) being
dependent on $\log m_{q,0}$ do not require any renormalization.
The main origin of differences of the non-leading logarithmic terms
in these formulae comes from the calculation procedure itself
since the integration over the emission gluon phase space
and tending of the quark mass to zero are not commutative with each other.

In order to find one-loop QCD contributions to the sum rules it is
necessary to integrate (\ref{qcdsf}) over the scaling variable $x$. After
an identical recombination like
\begin{equation}
\int \limits _0^1 dx\int \limits _x^1 dzR(z)f_q(x/z)=
\int \limits _0^1 dzzR(z)\int \limits _0^1 dxf_q(x)
\end{equation}
an explicit integration over $z$ gives
\begin{eqnarray}
\label{sr}
\int \limits _0^1 dx F_1(x,Q^2)=&&
\Bigl(1-\frac 23\frac {\alpha_s}{\pi}\Bigr)
\int \limits _0^1 dx F_1^0(x),
\nonumber \\
\int \limits _0^1 \frac {dx}{x} F_2(x,Q^2)= &&
\int \limits _0^1 \frac {dx}{x} F_2^0(x),
\nonumber \\
\int \limits _0^1 dx g_1(x,Q^2)=&&
\Bigl(1-\frac 53\frac {\alpha_s}{\pi}\Bigr)
\int \limits _0^1 dx g_1^0(x),
\nonumber \\
\int \limits _0^1 dx g_2(x,Q^2) =&& 0,
\end{eqnarray}
where the structure functions with index "0" are defined by
(\ref{0sf}).

Notice that QCD RC to the first moment to the unpolarized structure
functions as well as $g_2$ have identical values for both
massive and massless quark approaches. At the same time these two
methods give the different results for the QCD corrections to the
first moment of $g_1$.

Till present time the calculations have been performed within the
naive parton model when the quark mass is defined as in (\ref{mq},
\ref{m0}). Another possibility
is to consider it as a constant like it was assumed in \cite{AISh}. An
additional factor can depend only on $z$ and $x$. So the result for the
structure functions would differ only in an argument of logarithm
containing the quark mass. An additional contribution to the sum rules can
appear after integration over $x$ and $z$.  However it could be shown by
the explicit calculations that this contribution is equal to
zero. Indeed an
additional integrand has DGLAP-like structure, so this cancellation is
consequence of DGLAP equations. Thus the final results for sum
rules in
these two cases are identical.

\section{Discussion and Conclusion}

In this paper we applied the approach traditionally used in QED and
electroweak theory for calculation of QCD correction to the
DIS structure
functions and sum rules. Since the quark was considered massive
within this ap\-p\-ro\-ach, it allowed us to estimate the finite
quark mass effects
at the NLO level. LO correction  contributes to the DIS structure
functions but vanishes for the
sum rules, so this mass effects
are important just for the sum rules.

We found that there is no any additional effect for the first moment of
the unpolarized structure functions as well as for $g_2$. However there is
some non-zero correction to the first moment of $g_1$ and as a result
to the
Ellis-Jaffe and Bjorken sum rules. A value of the correction is in
agreement with the results of refs.\cite{NE2,TERV}, obtained by different
methods. We confirm also the statement of \cite{NE2} that the
classical
value of correction to the first moment of $g_1$ is reproduced if
we take into account the
leading term of the polarization vector $\eta=p_{1q}/m_q$. However
contrary to the paper we would interpret the result with (-5/3)
in eqs.(\ref{sr}) as
physical one. In our calculation we took the proton (quark)
polarization vector in a general form. Using of the exact
representation of the
vector \cite{ASh} ($k_1$ is an initial lepton momentum and
$k_1^2=m_l^2$):
\begin{equation}
\eta=\frac1{\sqrt{(k_1p_{1q})^2-m_l^2m_q^2}}
\Bigl[\frac {k_1p_{1q}}{m_q}p_{1q}-m_qk_1\Bigr]
\label{et}
\end{equation}
leads to exactly the same result (with -5/3) for the first
moment of $g_1$. The second term giving the additional
contribution (-2/3) cannot be neglected within the approximation
under consideration. It can be easyly understood from pure
calculation of QED and electroweak corrections to the lepton current
in DIS process \cite{LRC,BARD}, where the lepton polarization vector
looks like (\ref{et}), and this additional contribution was extracted to
separate non-vanishing contribution. The result with (-5/3) was
obtained under similar assumptions in the report \cite{TER}. The
additional correction was discussed both within OPE (operator
product expansion) and within improved quark-parton model in
paper \cite{TERV}. It was shown in this paper that there is no
contradiction between this result and classical correction obtained for
massless QCD, if we carefully take into account the finite
quark mass effects for coefficient function and matrix element.
As it was reviewed in recent paper \cite{NE3} the
renormalization of the axial-vector current completely
suppresses the additional contribution obtained within massive approach
(see section 4 of the paper).

We note that Burkhard-Cottingham sum rule is held withing taking
into account mass effects at the NLO level. The target mass
correction of the next order ($\sim m_q^2/Q^2$) was analysed and
was found negative in the paper \cite{TERM}. However, in the case
of conserved currents (see \cite{BT}) the polarized structure
function $g_2$ with the target mass correction obeys not only
Burkhard-Cottingham sum rule but Wandzura-Wilczek relation
too.

Estimating the diagrams with one gluon radiation we have also result
for QED radiative correction to had\-ro\-nic current. The calculation
within the quark parton model shows that QED correction is not so large,
however there are at least two arguments to be taken into account.
First, the DIS structure functions are defined in the one-photon
exchange approximation as an objects including only  strong
interaction. It means that transferring from the cross section to the
structure functions we neglect these QED effects, so their contributions
have to be included to systematic error of corresponding measurements.
Second reason is that there exist some measurements where  QED
correction is important. One of the examples is calculation of
$\alpha_s $ by comparing theoretical and experimental values for
the Bjorken sum rule \cite{EK}. In this case $\alpha_s$ is
suggested to be extracted from QCD corrections to the Bjorken sum
rule, which is a series over $\alpha_s $, and at least five terms
of the expansion are known. Simple estimation shows that QED
contribution is comparable with fourth (or even third)
term of the expansion.

There are several papers devoted to calculation of QED and electroweak
corrections to the hadronic current \cite{AISh,BARD,SP}. As usual methods
similar to ours are used for that. Positive moment here is the keeping
quark
mass non-zero. From the other side there are some effects which are not
considered within QED calculations: taking into account the confiment
effect in integration of soft region, consideration of the quarks as
non-free particles. Thus the methods (see for example \cite{TER})
developed for careful treatment of the QCD effects should be
applied for the analysis of photon emission from the hadronic
current. However, probably the best way to take into account the
photon
radiative correction to the hadronic current is further
generalization of OPE technique to include QED effects.

\section*{Acknowledgement}

We benefitted much for discussion with A.V. Efremov, E.A.Ku\-ra\-ev
 N.N. Nikolaev and O.V.Teryaev.
One of us (I. A.) thanks the US Department of Energy for
support under contract DE-AC05-84ER40150.

\appendix
\renewcommand{\thesection}{Appendix}
\renewcommand{\thesubsection}{\Alph{section}.\arabic{subsection}}
\renewcommand{\theequation}{\Alph{section}.\arabic{equation}}
\section{}
\setcounter{equation}{0}

The gluon emission  phase space
on the parton level can be written in a covariant form:
\begin{equation}
\frac {d^3k}{k_0}=
\frac{(Q^2+u_q-v_q)}{2(Q^2+u_q)}
\frac {du_qdv_qdz_1}{\sqrt {-\Delta (k_1, k_2, p_{1q}, p_{2q}})}.
\label{phs}
\end{equation}
Here $v_q=2kp_{1q}$, $u_q=2kp_{2q}$, $z_1=2kk_1$, $\Delta (k_1, k_2,
p_{1q}, p_{2q})$ is Gram determinant,
$\lambda _q=(Q^2+u_q)^2+4m_q^2Q^2$
 and $p_{1q}$ ($p_{2q}$) is an
initial (final) quark momentum.

Since the gluon is radiated from the quark legs it is clear that the
 matrix element squared for this process does not depend on $z_1$.
Therefore an analytical integration of the right part of (\ref{phs})
over $z_1$ can
 be performed
\begin{equation}
\int \limits _{z_1^{min}}^{z_1^{max}}\frac{dz_1}{2\sqrt {-\Delta (k_1,
k_2,
p_{1q}, p_{2q}})}
=\frac {\pi }{\sqrt{\lambda _q}},
\end{equation}
where the limits $z_1^{max/min}$ are defined as the solutions of
the equation
\begin{equation}
\Delta ( k_1, k_2, p_{1q}, p_{2q} )=0.
\end{equation}
The following expression
\begin{equation}
J[A]=\int \limits ^{v^{max}_q} _{v^{min}_q} dv_qA
\end{equation}
for the integration over $v_q$ will be used in
this appendix.
Explicit expressions for the limits of integration
which can be obtained as the solution of the equation
$z_1^{max}=z_1^{min}$ read:
\begin{equation}
v_q^{max/min}=\frac
{u_q(Q^2+u_q+2m_q^2\pm \sqrt {\lambda _q})}{2\tau},
\end{equation}
where $\tau=u_q+m_q^2 $.

Tensor and vector integrals can be reduced to scalar ones using the
following expressions:
\begin{eqnarray}
\displaystyle
J \Bigl[ k_\mu k_ \nu A \Bigl]=
&&
\frac {g_{\mu \nu }}{2\lambda _q}J\Bigl[
T_1A \Bigl]
+\frac {x^2 p_\mu p_\nu }{\lambda _q^2}J \Bigl [ T_2A\Bigl]
\nonumber\\&&
+\frac {x(p_\mu q_\nu + q_\mu p_\nu)}{\lambda _q^2} J \Bigl[ T_3A\Bigl ]
+\frac {q_\mu q_\nu }{\lambda _q^2}J \Bigl[ T_4A \Bigl ] ,
\nonumber \\
J \Bigl[ k_\mu A \Bigl ] =
&&
\frac {xp_\mu }{\lambda _q} J \Bigl[ V_2A \Bigl
]+\frac {q_\mu}{\lambda _q} J \Bigl[ V_2A \Bigl ].
\label{vect}
\end{eqnarray}
The coefficients $T_i$ and $V_i$
are defined as the solution of the
systems of equations  obtained by the convolutions
(\ref{vect}) with $p_{1q}$, $q$ and $g_{\mu \nu }$ :
\begin{eqnarray}
T_1=
&&
(u_q-v_q)^2m_q^2+u_qv_q(v_q-u_q-Q^2),
\nonumber \\
T_2=
&&
((u_q+v_q)^2+2u_qv_q)Q^4
+2(u_q+2v_q)Q^2u_q
\nonumber \\
&&
+(u_q^2-2m_q^2Q^2)(u_q-v_q)^2,
\nonumber \\
T_3=
&&
(4u_q-v_q)Q^2u_qv_q+(2u_q+v_q)Q^3v_q
\nonumber \\
&&
-m_q^2(u_q-v_q)(3u_q(Q^2+u_q-v_q)+Q^2v_q),
\nonumber \\
T_4=&&
(Q^2+u_q)^2v_q^2+6(u_q-v_q)^2m_q^4
\nonumber \\
&&
-2m_q^2v_q[(Q^2+u_q)(3u_q-2v_q)
-u_qv_q],
\nonumber \\
V_1=
&&
Q^2(u_q+v_q)+u_q(u_q-v_q),
\nonumber \\
V_2=
&&
v_q(u_q+Q^2)+2m_q^2(v_q-u_q).
\end{eqnarray}
The list of scalar integrals over variable $v_q$ in NLO
approximation reads as
\begin{eqnarray}
J[\frac 1 {v_q^2}]=&&\frac{\sqrt {\lambda _q}}{m_q^2u_q},
\nonumber \\
J[\frac 1 v_q]=&&\log \frac {(Q^2+u_q+2m_q^2+\sqrt {\lambda
_q})^2} {4m_q^2\tau},
\nonumber \\
J[1]=&&\frac{u_q\sqrt {\lambda _q}} {\tau},
\nonumber \\
J[v_q]=&&\frac{u_q^2(Q^2+u_q+2m_q^2)\sqrt {\lambda _q}} {2\tau
^2},
\end{eqnarray}

\end{document}